# Anomalous elastic buckling of hexagonal layered crystalline materials in the absence of structure slenderness

Manrui Ren[1], Jeffernson Zhe Liu[2*], Lifeng Wang[4], and Quanshui Zheng[1,3*]

[1] Department of Engineering Mechanics and Center for Nano and Micro Mechanics, Tsinghua University, Beijing 100084, China

[2] Department of Mechanical and Aerospace Engineering, Monash University, Clayton, VIC 3800, Australia

[3] Institute of Advanced Study, Nanchang University, Nanchang 330031, China

[4] Department of Mechanical Engineering, Stony Brook University, Stony Brook, NY 11794-2300, United States

**Abstract**

Hexagonal layered crystalline materials, such as graphene, boron nitride, tungsten sulfate, and so on, have attracted enormous attentions, due to their unique combination of atomistic structures and superior thermal, mechanical, and physical properties. Making use of mechanical buckling is a promising route to control their structural morphology and thus tune their physical properties, giving rise to many novel applications. In this paper, we employ finite element analysis (FEA), molecular dynamic (MD) simulations and continuum modeling to study the mechanical buckling of a column made of layered crystalline materials with the crystal layers parallel to the longitudinal axis. It is found that the mechanical buckling exhibits a gradual transition from a bending mode to a shear mode of instability with the reduction of slenderness ratio. As the slenderness ratio approaches to zero, the critical buckling strain $\varepsilon_{cr}$ converges to a finite value that is much lower than the material's mechanical strength, indicating that it is realizable under appropriate experimental conditions. Such a mechanical buckling mode is anomalous and counter-intuitive. Our continuum mechanics model for the critical bucking strain (at a zero slenderness ratio) agrees very well with the results from the FEA simulations for a group of typical hexagonal layered crystalline materials. MD simulations on graphite indicate the continuum mechanics model is applicable down to a scale of 20 nm. Our theoretical model reveals that the critical bucking strain (at a zero slenderness ratio) solely depends on the material's elastic constants (with no structural dimensions), implying that it is an intrinsic material property. A new concept, intrinsic buckling strain, is defined in this paper. This study provides avenues for engineering layered crystalline materials in various nano-materials and nano-devices via mechanical buckling.

# 1. Introduction

Buckling, as a mechanical instability, is a common phenomenon in nature (Gere and Timoshenko, 1998; Price and Cosgrove, 1990; Zartman and Shvartsman, 2010). It is often treated as a nuisance to be avoided. This view is changing with the growing knowledge of this phenomenon (Biot, 1957; Bowden et al., 1998; Brau et al., 2010; Budd et al., 2003; Efimenko et al., 2005; Gere and Timoshenko, 1998; Hohlfeld and Mahadevan, 2011; Huang et al., 2005; Hunt et al., 2000; P. Kim et al., 2011; Pocivavsek et al., 2008; Wadee et al., 2004) and the emerging successful cases of employing mechanical buckling in real applications (Efimenko et al., 2005; Guo et al., 2011; D. H. Kim et al., 2008; J. Kim et al., 2009; R. H. Kim et al., 2010; Koo et al., 2010; Rogers et al., 2010; Stafford et al., 2004; Y. Wang et al., 2011; Zang et al., 2013). For example, utilizing buckled interconnecting components in electronic devices leads to "stretchable electronics" that can accommodate large stretching and compressive loads without breaking (D. H. Kim et al., 2008; Rogers et al., 2010). Mechanical buckling of a thin stiff film on a soft substrate under an in-plane compression can alter the surface morphology and thus modulate the surface physicochemical properties, giving rise to various applications, such as artificial skins (Efimenko et al., 2005), micro-devices to measure mechanical properties of thin polymer and nanoparticle films (Leahy et al., 2010; Stafford et al., 2004), dynamically controlled surface wettability (Zang et al., 2013), enhancement of light extracting efficiency from organic light-emitting diodes (Koo et al., 2010), and dynamic display of biomolecule micropatterns (J. Kim et al., 2009). The surface ripples also have many applications in micro-fluidic devices (Efimenko et al., 2005) and artificial muscle actuators (Zang et al., 2013).

The discovery of graphene (Novoselov et al., 2004) has stimulated intensive research interests for two dimensional crystalline materials, such as BN, $MoS_2$, $WS_2$, silicene, graphyne, and so on (Golberg et al., 2010; Malko et al., 2012; Nicolosi et al., 2013; Vogt et al., 2012; Q. H. Wang et al., 2012; Wilson and Yoffe, 1969). For this class of materials, atoms are distributed in a layered crystal lattice and are bonded via strong chemical bonds, whereas different crystal layers interact with each other through weak van der Waals or electrostatic forces. Such a two dimensional crystalline material has a unique combination of structural, mechanical and physical properties, enabling great potentials for applications in electronic devices, catalysts, batteries, and super-capacitors, as seen in recent extensive experimental and theoretical studies (Geim and Novoselov, 2007). In practice, these materials are often fabricated in a form with multiple crystal layers stacked together, either for the convenience of fabrication or intentionally. For example, tuning either the number of layers or the stacking sequence of different types of crystal layers can modulate electronic properties of the resultant van der Waals heterostructures (Geim and Grigorieva, 2013; Haigh et al., 2012). It turns out that using multi-layers of graphene as a building block of graphene cellular foams is essential for the observed super-elasticity under a large compressive strain up to 80% in experiments (Qiu et al., 2012).

In addition to the widely studied approaches to tailor the physical properties of layered crystalline materials, *e.g.*, scissoring graphene into different shapes (Ci et al., 2008), chemical doping (Ci et al., 2010), chemical or physical adsorption (Elias et al., 2009; Nair et al., 2010; Schedin et al., 2007; Xu et al., 2009), mechanical buckling caused by a compressive load parallel to the basal planes can serve as a promising method to enable new applications. There are already several successful experimental studies. It has been reported that the reversible mechanical buckling of a stack of graphene-oxide layers is the origin for the hydration responsive property of graphene oxide liquid crystal in experiments (Guo et al., 2011). The periodically rippled graphene ribbons formed on a pre-stretched elastomer substrate can be used as high performance strain sensors (Y. Wang et al., 2011). A super-hydrophobic surface with a reversibly tunable wettability has been realized using crumpled graphene films (Zang et al., 2013). However, employing mechanical buckling of layered crystalline materials in applications is still hampered by inadequate understanding of this phenomenon.

The most well-known elastic buckling is the bending mode of instability studied back to Euler's era (Fig. 1(c)) (Gere and Timoshenko, 1998). For a slender structure, such as a beam, plate, or thin film, being subject to a longitudinal compression, lateral deflection will occur beyond a critical load. This is because bending is energetically less costly than compression for these slender structures. Most of the applications described previously are based on this type of instability. It should be noted that the unique atomistic structures of layered materials, *i.e.*, strong in-plane covalent chemical bonds, giving rise to a very high in-plane elastic modulus, and weak out-of-plane van der Waals or electrostatic interactions, yielding a very small interlayer shear modulus, implies a distinctive shear mode of instability (Fig. 1(b)). Under this mode, above a critical compressive load along in-plane directions, the shear deformation among adjacent atomic layers occurs, generating a lateral displacement and then releasing the compressive strain. Such a shear mode of instability was observed in wood (Byskov et al., 2002), fiber reinforce composites (Budiansky et al., 1998; Kyriakides et al., 1995), and geological strata (Price and Cosgrove, 1990). However, there are very few experimental and theoretical studies for the shear mode instability of the layered crystalline solids (Z. Liu et al., 2010), particularly in terms of the critical buckling load.

In this paper, we firstly use finite element analysis (FEA) to investigate the elastic bucking of a column made of the most well-known hexagonal layered crystalline material, graphite, under a compressive load parallel to the basal plane of its crystal layers (in section 2). With reduction of its slenderness ratio, the elastic buckling evolves from the bending mode of instability to the shear mode of instability. As the slenderness ratio approaches to zero, the critical strain of buckling converges to a small value (0.86%), in contrast with the classical Euler model. In section 3 of this paper, a continuum mechanics model is developed for the critical buckling strain at an infinitesimal slenderness ratio. FEA simulations for a group of hexagonal layered crystalline materials are used to verify this model. In section 4, molecular dynamic (MD) simulations are employed to simulate the mechanical buckling of graphite at a nanometer scale. It shows that the continuum mechanics model provides accurate predictions for critical buckling strain down to 20 nm. Section 5 discusses the implications of such elastic buckling in layered crystalline materials and its potential applications. Conclusions are drawn in section 6.

**2. Finite element models**

Fig. 1(b) and (c) depict a graphite column, where the basal planes of graphene layers are parallel to the column longitudinal axis (*x*-axis). Both vertical displacement (*x*-axis) and rotation of the cross-section plane at bottom end is fixed. The cross-section plane of top end is subject to a vertical downward displacement $\delta_x$ and its rotation is fixed. A constraint is applied to ensure that centers of the two cross-sections are on the same *x*-axis. Several typical cross-section shapes are considered, *e.g.*, circular, square, and triangles. A slenderness ratio is defined as a ratio of the column axial dimension *L* (along *x*-axis) over the gyration radius of the cross-section $\rho$. Fig. 1(c) shows a slender column with a large $L/\rho$ and Fig. 1(b) depicts a short column with a low value of $L/\rho$. The commercial FEA software ABAQUS is employed to determine the critical buckling loads of the graphite columns, using the BUCKLE module. Tetrahedral volume elements are used to mesh the three dimensional columns. A linear transverse isotropic constitutive relation for graphite is adopted (Kelly, 1981). The critical buckling strain $\varepsilon_{cr}$ is defined as a ratio of the top-end displacement at the elastic buckling point $\delta_{cr}$, which is determined using ABAQUS, over the column longitudinal dimension, *i.e.*, $\varepsilon_{cr} = \delta_{cr}/L$.

Fig. 1(a) shows the results of $\varepsilon_{cr}$ as a function of the slenderness ratio $L/\rho$ for three different cross-section geometries, *e.g.*, circular, square, and regular triangle. Other types of cross-sections, like isosceles triangles with different vertex angles and rectangular with different aspect ratios, are also examined. For a given $L/\rho$, the obtained $\varepsilon_{cr}$ results for different cross-section shapes almost overlap with each other (Fig. 1(a)), indicating that the cross-section geometry has a minor effect on the obtained relation of $\varepsilon_{cr}$ versus $L/\rho$. In Fig. 1(a), the critical strain $\varepsilon_{cr}$ increases with a reduction of $L/\rho$. For a large slenderness ratio (> 100), the FEA results agree with the classical Euler model very well, implying the nature of bending mode buckling. Indeed, from ABAQUS calculations, there is compressive strain on one side and expansive deformation on the opposite side of the buckled column ($L/\rho > 100$), which is consistent with the typical strain distribution of a beam under bending. In the Euler model, the $\varepsilon_{cr}$ for a column with both ends fixed (Fig. 1(c)) is

$$\varepsilon_{cr} = 4\pi^2 \left(\frac{L}{\rho}\right)^{-2} \tag{1}$$

However, a drastic difference is observed for a small or medium slenderness ratio ($L/\rho < 100$). It is interesting that as the $L/\rho$ approaches to zero, the FEA results converge to a small constant of ~ 0.86%, in contrast with infinity as predicted from the Euler theory (Eq. (1)). Note that this calculated critical strain is far smaller than the mechanical strength of graphene. Recent experiments showed that graphene could sustain an in-plane mechanical strain of up to 20-30% (C. Lee et al., 2008; G. H. Lee et al., 2013). In principle, elastic buckling should take place prior to mechanical failure for a graphite column under a compressive load, even in the case of $L/\rho \sim 0$. This is an anomalous mechanical buckling, contrary to the prediction of the classic Euler theory.

To understand the counter-intuitive results from ABAQUS simulations, strain distribution in the buckled column with a small $L/\rho$ is carefully studied. Our FEA numerical results show a profound shear strain $\varepsilon_{xz}$ throughout the column and a negligible normal strain $\varepsilon_x$, suggesting a shear mode of instability. It is well known that the Timoshenko beam theory has taken the shear effect into account. Fig. 1(a) also shows the $\varepsilon_{cr}$ results from Timoshenko theory for a column as depicted in Fig. 1(c)

$$\varepsilon_{cr} = 4\pi^2 \left[\left(\frac{L}{\rho}\right)^2 + 4\pi^2 \frac{nY}{G}\right]^{-1} \tag{2}$$

where $Y$ is Young's modulus of the column along longitudinal $x$ direction and $G$ is the shear modulus (Timoshenko and Gere, 2012). Here $n$ is a factor related to the geometry of the cross-section. It equals to 1.12 and 1.11 for rectangular and circular cross-sections, respectively. Timoshenko theory agrees with FEA results quite well for a medium slenderness ratio, *e.g.*, $70 < L/\rho < 100$. But there is a significant discrepancy for $L/\rho < \sim 70$. Taking the slenderness ratio approaching zero in Eq. (2) leads to a critical strain $\varepsilon_{cr} = G/nY$. For a graphite column with a rectangular cross-section, it is approximately 0.38%, only about one half of the FEA result. It is reasonable to see such a discrepancy, particularly at a low $L/\rho$, because Timoshenko beam theory incorporates a mixture of bending and shearing modes, implicitly assuming a relative large slenderness ratio.

In composite structures, such as fiber reinforced composites, wood, and geological strata, the shear mode of instability is often observed. In order to estimate the critical stress at which the fiber undergoes buckling, Rosen (Rosen 1965) modeled the fiber embedded in matrix as a beam embedded in an elastic foundation. It concluded a critical stress as

$$\sigma_c = \frac{G_m}{(1-v_f)} \tag{3}$$

where $G_m$ is the shear modulus of matrix and $v_f$ is the volume fraction of fibers. For a graphite column, using the rule of mixing leads to $\sigma_c = G$, the shear modulus of the composite. Thus the critical strain can be estimated as $\varepsilon_{cr} = G/Y$, which is very close to that of Timoshenko beam theory. Clearly, neither Timoshenko beam theory nor the Rosen model can describe the calculated the critical load of elastic buckling of a graphite column in FEA (Fig. 1(a)).

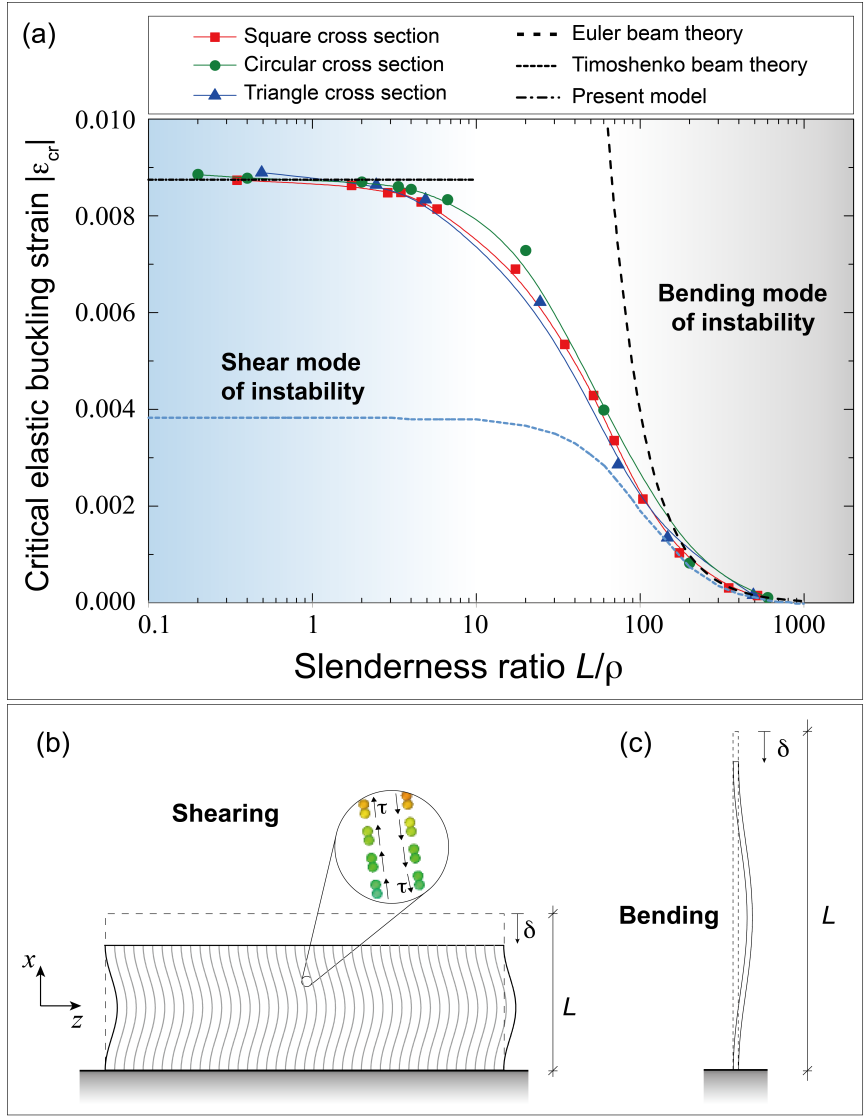

**Fig. 1**. (a) Critical elastic buckling strain $\varepsilon_{cr}$ of a column made of the most well-known hexagonal layered crystalline material, graphite, calculated using the finite element analysis (FEA). (b) and (c) Graphite columns with a small and large slenderness ratio, respectively. Basal planes of graphite are parallel to the column longitudinal axis. Both ends of the column are (clamp-) fixed and a compressive displacement is applied along the longitudinal direction. Different slenderness ratio $L/\rho$ is examined, where $L$ is the length of the column and $\rho$ is the radius of gyration of the cross-section. Results for three different cross-section shapes, *e.g.*, circular, square, and regular triangle, are shown in (a). Excellent agreement with the Euler theory for those columns with a large $L/\rho$ (>100) indicates the nature of bending mode instability. The short columns ($L/\rho$ < 10) exhibit a shear mode of instability.

## 3. Continuum mechanics model

In this section, a theoretical model is developed for the critical elastic buckling strain of a column made of a hexagonal layered crystalline material with a small slenderness ratio. Fig. 1(a) shows that the critical strain $\varepsilon_{cr}$ is nearly a constant in the case of $L/\rho < 10$. For simplicity, the extreme case of $L/\rho = 0$ is considered. Fig. 1(b) depicts our column model and established coordinate system. Basal planes of the crystal layers are parallel to the *x-y* plane. The origin point is placed in the center of the column and the *y-z* plane overlaps with the middle cross-section plane. A periodic boundary condition is applied in the lateral direction (*z*-axis), yielding $\rho \rightarrow \infty$ and thus $L/\rho = 0$. This system is modeled as a plane-strain problem (in the *x-z* plane), aiming to be consistent with some popular experimental setups, *i.e.*, utilizing strain mismatch between a film made of layered crystalline materials and a pre-stretched substrate to drive the elastic buckling (Bowden et al., 1998; Efimenko et al., 2005; P. Kim et al., 2011; Koo et al., 2010; Zang et al., 2013). Given that the *z*-axis in our model is the vertical direction of the film/substrate system in experiments, a compressive strain (via release of the pre-stretch deformation of substrate) is applied along the basal plane direction (*x*-axis), meanwhile the vertical direction (*z*-axis) is free to relax. Under the constraint from the substrate, deformation in *y* direction is negligible, leading to the plane-strain condition.

For the column model shown in Fig. 1(b), boundary conditions are

$$u_1\big|_{x=L/2} = -\delta/2, \quad \tau_{xy}\big|_{x=L/2} = 0, \quad \tau_{xz}\big|_{x=L/2} = 0$$
$$u_1\big|_{x=-L/2} = \delta/2, \quad \tau_{xy}\big|_{x=-L/2} = 0, \quad \tau_{xz}\big|_{x=-L/2} = 0 \quad (4)$$
$$u_3\big|_{x=-L/2} = u_3\big|_{x=L/2}$$

where $\delta$ is the relative displacement of the both ends moving toward each other. Note that boundary conditions for displacements are consistent with those of the FEA models, in which both ends are fixed. The displacement fields can be expressed as

$$u_1 = \varepsilon x + \sum_{n=1}^{N} a_n \frac{L}{2n\pi} \sin\frac{2n\pi x}{L}, \quad u_2 = 0, \quad u_3 = -\varepsilon d_{13} z + \sum_{n=1}^{N} b_n \frac{L}{2n\pi} \cos\frac{2n\pi x}{L} \quad (5)$$

where $\varepsilon = \delta/L$ represent the homogeneous compressive strain in *x*-direction prior to elastic buckling and $d_{13} = C_{13}/C_{33}$, in which $C_{13}$ and $C_{33}$ are elastic constants. Clearly, $u_1$ is an odd function of coordinate *x*. Thus, a series composed of sine functions is used to represent the displacement after the elastic buckling. Note that $u_1$ is independent of *z* because of the periodic boundary condition in *z* direction. The displacement $u_3$ is an even function of coordinate *x*. It is thus expressed as a series made of cosine functions. The first term of $u_3$ is adopted for a purpose of releasing normal stress $\sigma_z$ upon the compressive load $\delta$ in $x$ direction. These displacement fields (Eq. (5)) satisfy the boundary conditions (Eq. (4)). Parameters $a_n$ and $b_n$ are unknown coefficients. Only when the load is above the critical buckling value $\delta_{cr}$, the $a_n$ and $b_n$ will have nonzero solutions.

Deformation gradient $\boldsymbol{F}$ and the first Seth strain $\boldsymbol{E}$ can be derived based on the displacement fields as

$$F_{11} = 1 + \frac{\partial u_1}{\partial x}, \quad F_{33} = 1 + \frac{\partial u_3}{\partial z}, \quad F_{13} = \frac{\partial u_1}{\partial z}, \quad F_{31} = \frac{\partial u_3}{\partial x} \tag{6}$$

$$E_{11} = \left(F_{11}^2 + F_{31}^2 - 1\right)/2, \quad E_{33} = \left(F_{33}^2 + F_{13}^2 - 1\right)/2, \quad E_{13} = (F_{11}F_{13} + F_{31}F_{33})/2 \tag{7}$$

A hexagonal layered crystalline material has a transversely isotropic elasticity. In our model, since its basal plane is in the $x$-$y$ plane, the constitutive law of linear elasticity can be expressed as

$$\begin{pmatrix} \sigma_x \\ \sigma_y \\ \sigma_z \\ \tau_{yz} \\ \tau_{zx} \\ \tau_{xy} \end{pmatrix} = \begin{pmatrix} C_{11} & C_{12} & C_{13} & & & \\ C_{12} & C_{11} & C_{13} & & & \\ C_{13} & C_{13} & C_{33} & & & \\ & & & C_{44} & & \\ & & & & C_{44} & \\ & & & & & (C_{11}-C_{12})/2 \end{pmatrix} \begin{pmatrix} \varepsilon_x \\ \varepsilon_y \\ \varepsilon_z \\ \gamma_{yz} \\ \gamma_{zx} \\ \gamma_{xy} \end{pmatrix} \tag{8}$$

where $\sigma$ and $\tau$ represent the normal and shear stress components, $\varepsilon$ and $\gamma$ denote the normal and shear strain components, and $C_{ij}$ are the stiffness constants. Substituting Eq. (7) into Eq. (8) yields the second Piola-Kirchhoff stress $\boldsymbol{T}$ as

$$T_{11} = C_{11}E_{11} + C_{13}E_{33}, \quad T_{33} = C_{13}E_{11} + C_{33}E_{33}, \quad T_{13} = 2C_{44}E_{13} \tag{9}$$

Then the strain energy density $U$ is

$$U = 1/2(T_{11}E_{11} + 4T_{13}E_{13} + T_{33}E_{33}) \tag{10}$$

Integrating the energy density $U$ in the column leads to the potential energy as

$$W = \frac{1}{L} \int_{-L/2}^{L/2} U dx. \tag{11}$$

in which a unit length is taken along the $y$ and $z$ directions, owing to the plane strain condition and the periodic boundary condition, respectively.

Following the principle of minimum total potential energy, partial derivatives of $W$ with respect to the undetermined coefficients $a_n$ and $b_n$ should be equal to zero. Thus,

$$g_1 = \partial W / \partial b_1 = b_1\left(\left(C_{11} - 4C_{44}d_{13} - C_{13}d_{13}\right)\varepsilon + 2C_{44}\right) + O\left(b_1\left(C_{11} + C_{13}d_{13}^2 + 4C_{44}d_{13}^2\right)\varepsilon^2 / 2\right) = 0$$

$$g_2 = \partial W / \partial b_2 = b_2\left(\left(C_{11} - 4C_{44}d_{13} - C_{13}d_{13}\right)\varepsilon + 2C_{44}\right) + O\left(b_2\left(C_{11} + C_{13}d_{13}^2 + 4C_{44}d_{13}^2\right)\varepsilon^2 / 2\right) = 0$$

$$g_3 = \partial W / \partial b_3 = b_3\left(\left(C_{11} - 4C_{44}d_{13} - C_{13}d_{13}\right)\varepsilon + 2C_{44}\right) + O\left(b_3\left(C_{11} + C_{13}d_{13}^2 + 4C_{44}d_{13}^2\right)\varepsilon^2 / 2\right) = 0$$

$$g_4 = \partial W / \partial b_4 = b_4\left(\left(C_{11} - 4C_{44}d_{13} - C_{13}d_{13}\right)\varepsilon + 2C_{44}\right) + O\left(b_4\left(C_{11} + C_{13}d_{13}^2 + 4C_{44}d_{13}^2\right)\varepsilon^2 / 2\right) = 0$$

......

$$f_1 = \partial W / \partial a_1 = a_1\left(\left(3C_{11} - d_{13}C_{13}\right)\varepsilon + C_{11}\right) + O\left(a_1\left(3C_{11} + C_{13}d_{13}^2\right)\varepsilon^2 / 2\right) = 0$$

$$f_2 = \partial W / \partial a_2 = a_2\left(\left(3C_{11} - d_{13}C_{13}\right)\varepsilon + C_{11}\right) + O\left(a_2\left(3C_{11} + C_{13}d_{13}^2\right)\varepsilon^2 / 2\right) = 0$$

$$f_3 = \partial W / \partial a_3 = a_3\left(\left(3C_{11} - d_{13}C_{13}\right)\varepsilon + C_{11}\right) + O\left(a_3\left(3C_{11} + C_{13}d_{13}^2\right)\varepsilon^2 / 2\right) = 0$$

$$f_4 = \partial W / \partial a_4 = a_4\left(\left(3C_{11} - d_{13}C_{13}\right)\varepsilon + C_{11}\right) + O\left(a_4\left(3C_{11} + C_{13}d_{13}^2\right)\varepsilon^2 / 2\right) = 0$$

...... (12)

where the quadratic and higher order terms of strain $\varepsilon$ are omitted. Note that derivatives with respect to $b_n$ ($n = 1, 2, 3, 4 \ldots$) always yields the same equation as

$$\left(C_{11} - 4C_{44}d_{13} - C_{13}d_{13}\right)\varepsilon + 2C_{44} = 0 \tag{13}$$

The derivatives with respect to $a_n$ ($n = 1, 2, 3, 4 \ldots$) leads to another equation as

$$\left(3C_{11} - d_{13}C_{13}\right)\varepsilon + C_{11} = 0 \tag{14}$$

From Eq. (13) or (14), we obtained the critical strain as

$$\varepsilon_{cr} = -\frac{2C_{44}}{C_{11} - \left(4C_{44} + C_{13}\right)d_{13}}, \tag{15}$$

or

$$\varepsilon_{cr} = -\frac{C_{11}}{3C_{11} - C_{13}d_{13}}. \tag{16}$$

Note that Eq. (15) was firstly reported in PhD thesis of one of the authors (J. Z. Liu, 2002).

The prediction from Eq. (15) or (16) is valid only if the magnitude of $\varepsilon_{cr}$ is small. Otherwise the higher order terms cannot be neglected in Eq. (12). From the aspect of physics, the adopted linear elasticity model may not be valid in the case of finite deformation. More importantly, only when the predicted $\varepsilon_{cr}$ is lower than the material's mechanical strength, the mechanical buckling could take place. Since Eq. (16) predicts a critical buckling strain $|\varepsilon_{cr}| > 1/3$, its prediction should not be considered reliable.

It is interesting to notice that the Eq. (15) only includes the elastic constants of materials without any structural dimensions, which is different form the Euler (Eq. (1)) and Timoshenko theory (Eq. (2)), suggesting that such a mechanical buckling is an intrinsic property of materials. It is reasonable to understand this feature because in the case of $L/\rho = 0$, the column (Fig. 1(b)) is inherently "structure-less". Here we define the critical buckling strain $\varepsilon_{cr}$ at $L/\rho = 0$ (*e.g.*, Eq. (15)) as the intrinsic buckling strain (IBS).

The parameter $n$ in displacement fields (Eq. (5)) represents different buckling modes. Interestingly, all the buckling modes share one degenerate eigenvalue, *i.e.*, the IBS $\varepsilon_{cr}$ in Eq. (15). It is well known that the first buckling mode of a column with a length of $L$, in principle, should be equivalent to the second buckling mode of a column with a double length of $2L$. Thus, with the reduction of $L$, the critical load of the first buckling mode of a column usually increases, as seen in the Euler model (Eq. (1)). Because of the degenerated single eigenvalue for all the buckling modes in our model, it is reasonable to understand that the $\varepsilon_{cr}$ results obtained from FEA simulations reach a plateau in the range of a small slenderness ratio (Fig. 1(a)).

Substituting the elastic constants of graphite (Kelly, 1981), $C_{11} = 1060$ GPa, $C_{12} = 180$ GPa, $C_{13} = 15$ GPa, $C_{44} = 4.5$ GPa, and $C_{33} = 36.5$ GPa into Eq. (15), we have the IBS $\varepsilon_{cr} = -0.0086$. It agrees with the FEA results for graphite (Fig. 1(a)) very well. To further verify our theoretical model, a list of hexagonal layered crystalline materials is examined. These materials are selected based on a thorough survey done by Wang and Zheng. for hexagonal crystal materials with an extreme elastic anisotropy degree (L.-F. Wang and Zheng, 2007). Fig. 2 and Table I compare the FEA results with the theoretical predictions from Eq. (15) or Eq. (16). Overall, the agreement is very good. Owing to the intrinsic layered atomic structures, most of the hexagonal layered crystalline materials have an in-plane elastic constant $C_{11}$ much larger than other elastic constants, particularly the shear modulus $C_{44}$. Thus, Eq. (15) often yields a smaller value of $\varepsilon_{cr}$ than that of Eq. (16). Only one exception in the materials that we visited, *i.e.*, InSe, for which Eq. (16) leads to a smaller value. However, this predicted $\varepsilon_{cr}$ appears to be much higher than its material's yield strain and thus the elastic buckling is practically impossible. It is listed here for a theoretical interest. The relative large discrepancy for InSe (in Table I) could be attributed to the omitted higher order strain terms in Eq. (12).

In Table I, for those layered materials that attract enormous attentions at present, such as graphite, h-BN, $MoS_2$, and $WS_2$, their IBS $\varepsilon_{cr}$ results are smaller than 15%. It is thus feasible to manipulate morphologies of these layered materials via the shear mode of elastic buckling and thus tune their physical properties in experiments for novel applications. More discussions will be provided later in this paper.

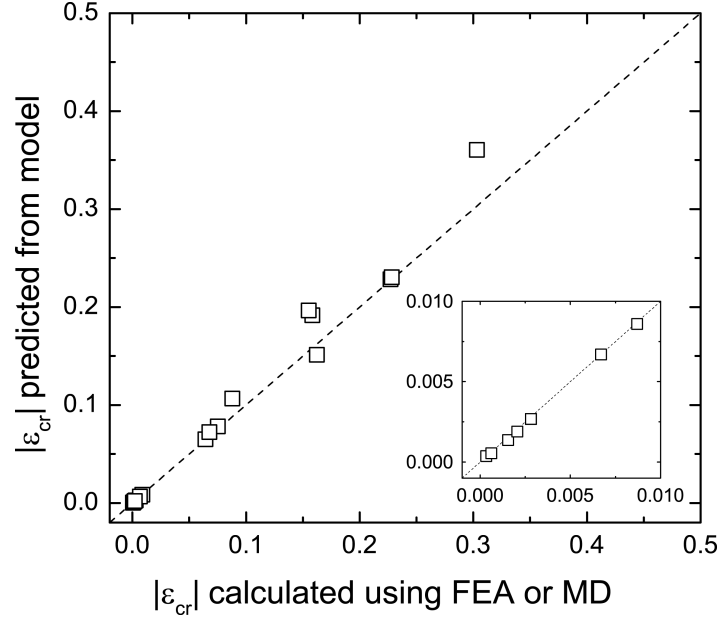

**Fig. 2**. A comparison of the critical elastic buckling strain results for a column with a slenderness ratio $L/\rho \sim 0$ (*i.e.*, the IBS $\varepsilon_{cr}$) predicted by using theoretical model (Eq. (15) or Eq. (16)) and those determined using finite element analysis (FEA) or molecular dynamic (MD) simulations. The column is made of hexagonal layered crystalline materials with a high degree of elastic anisotropy. Each symbol represents one type of hexagonal layered crystalline materials. Details of those materials and their IBS $\varepsilon_{cr}$ results are summarized in Table I and II.

**Table I**. Elastic constants, theoretical predictions (Eq. (15) or Eq. (16)) and FEA results of the IBS $\varepsilon_{cr}$ of selected hexagonal layered crystalline materials with a high degree of elastic anisotropy. Most of the materials are adopted from Table I in reference (L.-F. Wang and Zheng, 2007). Elastic constants of h-BN are from reference (Duclaux et al., 1992). Elastic constants of $WS_2$ are from reference (Volkova et al., 2012).

| Materials | $C_{11}$ | $C_{12}$ | $C_{13}$ | $C_{33}$ | $C_{44}$ | Theoretical prediction | FEA results |
|---|---|---|---|---|---|---|---|
| Graphite (C) | 1060 | 180 | 15 | 36.5 | 4.5 | −0.0086 | −0.0087 |
| Molybdenum sulfide ($MoS_2$) | 238 | −54 | 23 | 51 | 18.6 | −0.1917 | −0.1583 |
| Biotite [$K(Mg,Fe)_3AlSi_3O_{10}(OH,F)_2$] | 186 | 32 | 12 | 54 | 5.8 | −0.0651 | −0.0643 |
| Phlogopite [$KMg_3AlSi_3O_{10}(OH,F)_2$], B | 178 | 30 | 15 | 51 | 6.5 | −0.0783 | −0.0752 |
| Phlogopite [$KMg_3AlSi_3O_{10}(OH,F)_2$], A | 179 | 32 | 26 | 51.7 | 5.6 | −0.0724 | −0.0677 |
| Muscovite [$KAl_2Si_3O_{10}(OH,F)_2$] | 178 | 42.4 | 14.5 | 54.9 | 12.2 | −0.1513 | −0.1622 |
| Gallium sulfide (peizoel) (GaS) | 126.5 | 35.7 | 14.3 | 41.6 | 12 | −0.2284 | −0.2272 |
| Gallium selenide (peizoel) (GaSe) | 106.4 | 30 | 12.1 | 35.8 | 10.2 | −0.2305 | −0.2283 |
| Rubidium nickel chloride (RbNiCl$_3$) | 35.2 | 10.0 | 22 | 72.2 | 2.5 | −0.1965 | −0.1552 |
| Indium selenide (InSe) | 118.1 | 47.5 | 32 | 38.2 | 11.7 | −0.3606 | −0.3031 |
| Hexagonal Boron Nitride (h-BN) | 750 | 150 | – | 18.7 | 2.52 | −0.0067 | −0.0067 |
| Tungsten sulfide (WS$_2$) | 236 | 61 | 8 | 42 | 12 | −0.1065 | −0.0880 |

## 4. MD simulations

In light of great potentials of utilizing mechanical buckling in nano devices (Koo et al., 2010; Rogers et al., 2010; Zang et al., 2013), it is of great interests to investigate this phenomenon at a nanometer scale for the hexagonal layered crystalline materials. In this section, non-equilibrium molecular dynamic (NEMD) simulations is employed to study the elastic buckling of some graphite and virtual-graphite columns with a length $L$ down to 2 nm.

Fig. 3(a) depicts our molecular system: a graphite column composed of periodic A/B stacked graphene layers (along the $z$-axis) with their basal planes parallel to the longitudinal axis ($x$-direction). Longitudinal length $L$ is selected between 2 nm and 40 nm. Periodic boundary conditions (PBC) are applied in all three directions. The dashed box in Fig. 3(a) represents the super-cell used in our MD simulations. A constant velocity of $10^{-5} - 1$ Å/ps is applied to reduce the size of the super-cell in $x$ direction, meanwhile the $y$ and $z$ dimensions of the super-cell are fixed and the three super-cell vectors remain perpendicular to each other. The PBC along the $x$-axis restricts wavelength of the first-order buckling as $L$. The PBC in $y$ and $z$ directions result in an infinite radius of gyration $\rho$ of the cross-section (*i.e.*, $\rho \sim \infty$). Thus our molecular models have a ratio of slenderness $L/\rho = 0$. Note that these applied boundary conditions are consistent with the fixed end conditions in our theoretical and FEA models (Fig. 1(b) and (c)), except the boundary condition in $z$ direction. Care should be taken when comparing results of MD simulations with those from the theoretical model, which will be presented later.

NEMD simulations are performed using the LAMMPS code (Plimpton, 1995). The adaptive intermolecular reactive empirical bond order (AIREBO) potential (Stuart et al., 2000) is adopted to describe the interatomic interactions of the graphite column. Temperature of the whole system is fixed at 0.1 K using the Berendsen thermostat, with the temperature calculated after removing the center-of-mass velocity. A time step of 1.0 fs is used and the simulations are continued until the elastic buckling took place.

For a column with $L$ = 4.6 nm, Fig. 3(b) shows the results of potential energy and stress $\sigma_x$ as functions of compressive strain $\varepsilon_x$ in $x$ direction, at two different loading velocities $10^{-4}$ Å/ps and $10^{-2}$ Å/ps. The strain is defined as $\varepsilon_x = \delta/L$, where $\delta$ is the change of the unit cell in $x$-dimension. The results of potential energy and the stress in $x$-direction can be directly output from LAMMPS. In the beginning, both potential energy results appear as a parabolic function of the strain $\varepsilon_x$. This is consistent to the obtained linear stress-strain relations in Fig. 3(b). At the critical point, the energy curves start to deviate from the parabolic relation and accordingly the stress curves exhibit a significantly drop, indicating the happening of mechanical buckling. Indeed, the carbon atoms exhibit a clear lateral displacement in $z$ direction after the critical point. Color-map in Fig. 3(a) shows the relative magnitude of displacement obtained at a loading velocity of $10^{-4}$ Å/ps, in which the blue color denotes a relatively larger displacement than the red color. Clearly, this displacement profile qualitatively agrees with the FEA results. In Fig. 3(b), the determined critical strain value $\varepsilon_{cr}$ sensitively depends on the loading velocity. At $10^{-2}$ Å/ps, the $|\varepsilon_{cr}|$ equals to 0.2496%, whereas at a lower loading velocity of $10^{-4}$ Å/ps the critical strain result significantly reduces to $|\varepsilon_{cr}|$ = 0.1237%. This is a common phenomenon in dynamic buckling (Lindberg, 2003). After release of the compressive load, the buckled graphite column bounces back, fully recovering its original shape.

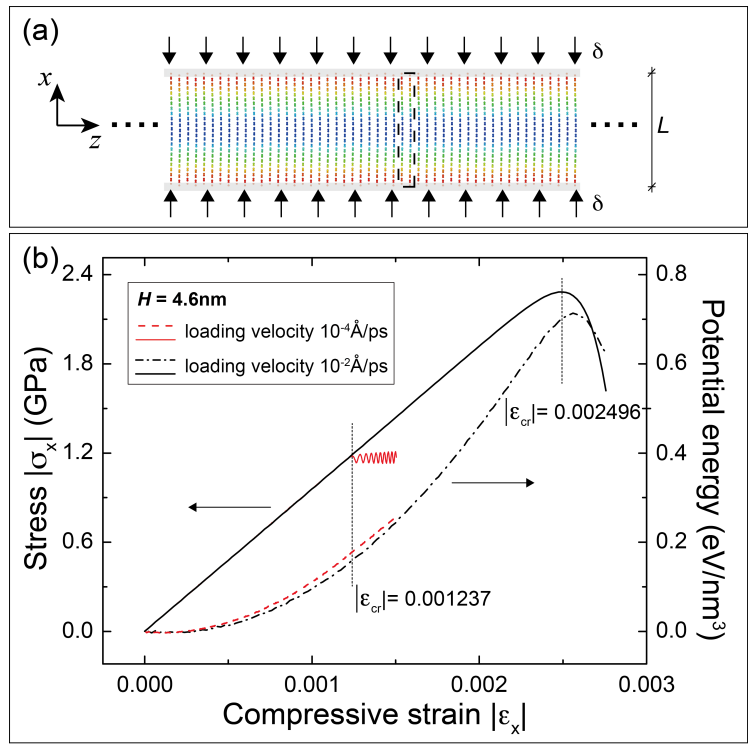

**Fig. 3**. (a) Molecular model for mechanical buckling of a graphite column in NEMD simulations. Graphene basal planes are parallel to the *x-y* plane. The box enclosed by dashed lines represents the super-cell. Periodic boundary conditions are applied in all three directions. Note that the periodic boundary conditions applied in lateral directions imply an infinite gyration radius $\rho$ and thus lead to a zero slenderness ratio $L/\rho$. A constant velocity of $10^{-5} - 1$ Å/ps is applied to reduce the size of the unit cell in *x* direction, meanwhile the *y* and *z* dimensions of the super-cell are fixed and the three super-cell vectors remain perpendicular to each other. The color map illustrates relative magnitude of the lateral displacement (*z*) after mechanical buckling, in which the blue color denotes a larger displacement than the red color. (b) Potential energy and stress $\sigma_x$ versus compressive strain $\varepsilon$ in MD simulations. At the critical point, the potential energy and stress results show an abrupt change, indicating the happening of mechanical buckling.

Fig. 4 summarizes the dependence of $|\varepsilon_{cr}|$ on the loading velocity for a graphite column with different length *L* from about 2 nm to 40 nm. Reduction of loading velocity generally leads to a decrease of obtained $\varepsilon_{cr}$ value. Apparently, a convergence is achieved below a loading velocity of $10^{-4}$ Å/ps. It is also found that a shorter column has a higher converged $\varepsilon_{cr}$ result. Above $L \sim 20$ nm, only a minor difference is observed in the converged $\varepsilon_{cr}$, *i.e.*, between –0.038% and –0.034%. The converged values should represent the IBS $\varepsilon_{cr}$ under a quasi-static condition, which is comparable with the FEA results and our theoretical model. However, the MD results are drastically lower than the FEA and theoretical results, *i.e.*, ~ –0.034% *vs.* ~ –0.86%. Such a huge discrepancy can be attributed to the shear modulus $C_{44}$ predicted from the AIREBO force field, which is far lower than experimental result (Kelly, 1981) that is used in our FEA simulations and theoretical model.

To determine the shear modulus $C_{44}$ of graphite described using the AIREBO force field, a simple shear deformation is applied to a 11-layers A/B stacked graphenes (with periodic boundary conditions along the two directions of basal planes). That is, each graphene layer is kept rigid and displaced with respect to each other in a direction parallel to the graphene basal plane. The shear displacement is a linear function of the layer's position in the stack. Coordinates of the atoms in the deformed graphite are then fed into an in-house FORTRAN code that can calculate Lennard-Jones (LJ) potential energy (to describe the van der Waals interactions among different graphene layers) in the AIREBO force field with the same cut-off distance used in the LAMMPS simulations. Magnitude of shear strain is selected to ensure the deformation within an elastic region, $-0.003 < \gamma < 0.003$. The obtained LJ potential energy of the graphene layer in middle of the stack exhibits a nearly perfect parabolic relation with respect to $\gamma$. Fitting the results using $U = 1/2C_{44}\gamma^2$ yields the shear modulus $C_{44}$ = 0.1783 GPa, which is significantly smaller than the experimental result 4.5 GPa. It is not a surprise to see such a large discrepancy. In the AIREBO force field model, two parameters $\varepsilon$ and $\sigma$ in the LJ potential $U_{LJ}(r) = 4\varepsilon[(\sigma/r)^{12}-(\sigma/r)^6]$ were fitted to reproduce two experimental results: interlayer distance 3.4 Å and elastic modulus $C_{33}$ = 36.5 GPa in $z$ direction. Therefore, the obtained Lennard-Jones model often provides unsatisfactory predictions of other physical properties, *e.g.*, binding energy between graphene layers and cleavage energy of graphite (Gould et al., 2013; Lebègue et al., 2010; Z. Liu et al., 2012; Sorella et al., 2009), and the interlayer shear modulus $C_{44}$.

In order to compare with our theoretical model, it should be aware that the boundary condition adopted in NEMD simulations in $z$-direction is different from that in the theoretical model. Since the $z$ dimension is fixed in NEMD simulations, the first term of $u_3$ in Eq. (5) should be zero. Setting $d_{13} = 0$ in Eq. (15) yields the IBS $\varepsilon_{cr} = -2C_{44}/C_{11} = -2\times 0.1783/980 = -0.03639\%$, in which the $C_{11}$ modulus is determined by fitting the potential energy curve as a fucntion of strain $U = 1/2C_{11}\varepsilon_x^2$ piror to the buckling (Fig. 3(b)). It agrees with the NEMD simulation results very well (Fig. 4) for $L \geq 19.68$ nm. The agreement is decent for $L = 9.63$ nm. But for a short graphite column with $L = 2.09$ nm or 4.61 nm, the difference is quite significant. It can conclude that our theoretical model provides accurate predictions of IBS $\varepsilon_{cr}$ down to a length scale of about 20 nm.

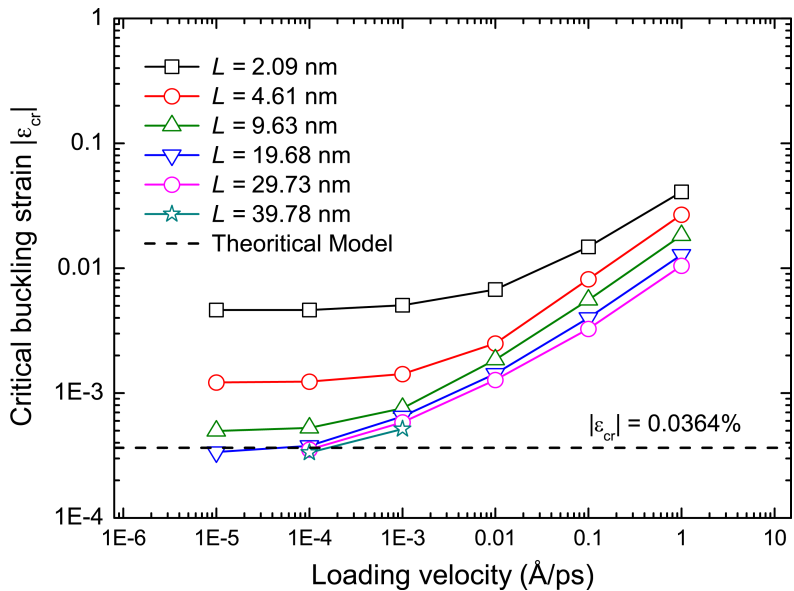

**Fig. 4**. Critical elastic buckling strain $\varepsilon_{cr}$ of a graphite column (Fig. 3(a)) as a function of loading velocity in NEMD simulations. Results for different longitudinal length $L$, *e.g.*, from 2.09 nm to 39.79 nm, are shown.

Our theoretical result for the IBS $\varepsilon_{cr}$ (dashed line) is shown for a comparison.

For a further verification, some *virtual* graphite models are created by artificially changing the parameter $\varepsilon$ in the LJ potential model with $\varepsilon = 0.00284$ eV and $\sigma = 0.34$ nm (Stuart et al., 2000). Given that the predicted $C_{44}$ value from the AIREBO is far smaller than the experimental results, the value of $\varepsilon$ is increased from 2 up to 10 times. In our MD simulations, the carbon-carbon interaction within one graphene layer is described using the Tersoff potential (Tersoff, 1989). Following similar procedures presented previously, for a column of length $L = 19.69$ nm, the critical buckling strain is determined in MD simulations using LAMMPS code at a loading velocity of $10^{-5}$ Å/ps. The shear modulus $C_{44}$ is calculated using the in-house FORTRAN code. The modulus $C_{11}$ is determined by fitting obtained potential energy as a fucntion of compressive strain prior to the buckling in MD simulations. The obtained IBS $\varepsilon_{cr}$ results are summarized in Table II and Fig. 2. The good agreement between the NEMD and the theoretical results, as seen in Table II, confirms the conclusion that our theoretical model for IBS $\varepsilon_{cr}$ is applicable down to a nanometer scale.

**TABLE II**. Comparison of IBS $\varepsilon_{cr}$ results from our theoretical model (Eq. (15)) and NEMD simulations. The elastic moduli $C_{11}$ and $C_{44}$ are shown as well.

| L-J parameter | $C_{11}$ (GPa) | $C_{44}$ (GPa) | Theoretical model | MD simulations |
|---|---|---|---|---|
| Graphite [a] | 980 | 0.1783 | −0.0003639 | −0.0003343 |
| Graphite ($\varepsilon, \sigma$) [b] | 1221 | 0.1727 | −0.0002829 | −0.0003480 |
| Virtual Graphite ($2\varepsilon, \sigma$) [b] | 1248 | 0.3462 | −0.0005546 | −0.0006110 |
| Virtual Graphite ($5\varepsilon, \sigma$) [b] | 1273 | 0.8704 | −0.001367 | −0.001425 |
| Virtual Graphite ($7\varepsilon, \sigma$) [b] | 1288 | 1.223 | −0.001899 | −0.001899 |
| Virtual Graphite ($10\varepsilon, \sigma$) [b] | 1309 | 1.758 | −0.002686 | −0.002656 |

[a] modeled by AIREBO force field model, the MD simulation results are obtained with $L = 39.79$ nm and loading velocity of $10^{-4}$ Å/ps.

[b] modeled by the Tersoff and the LJ potential models. The value of parameter $\varepsilon$ in LJ potential is increased by 2, 5, 7, and 10 times, respectively. Before the NEMD simulations and the calculations for $C_{44}$, interlayer distances and C-C bond lengths are optimized to reduce residual stresses. The MD simulation results are obtained with $L = 19.69$ nm and a loading velocity of $10^{-5}$ Å/ps.

## 5. Discussions

Before we draw conclusions, this section will provide some in-depth discussions and implications.

With a decrease of the slenderness ratio of a graphite column, the instability mode evolves from bending to shearing (Fig. 1(a)). Compared to isotropic materials, the shear mode is much more profound in layered hexagonal crystalline materials. The classical Euler model, which takes the bending deformation into account, only works for a graphite column with a very large slenderness ratio ($L/\rho > 100$). For a column with medium slenderness ratio ($10 < L/\rho < 100$), which is traditionally treated as a slender beam, the shear mode appears to play a significant role (Fig. 1(a)). The Timoshenko theory, which embodies a mixture of bending and shear deformation, works well in the rage of $70 < L/\rho < 100$. Our model represents a pure shear mode of deformation. The obtained IBS $\varepsilon_{cr}$ can successfully describe the critical buckling point of graphite columns with a slenderness ratio $L/\rho$ between 0 and 10. But a gap still exists. There are no appropriate models for $\varepsilon_{cr}$ at $10 < L/\rho < 70$ in Fig. 1(a).

It is worth noting that our continuum mechanical model embodies neither an intrinsic length scale nor internal atomistic microstructures. The graphite column is simply described as a homogeneous continuum bulk. It is interesting that the theoretical results of IBS $\varepsilon_{cr}$ agree very well with those determined by MD simulations (down to 20 nm). This excellent agreement suggests that the critical buckling point of graphite at a nanometer scale is governed by its macroscopic elastic properties. This should be true for other hexagonal layered crystalline materials (Table I). This conclusion is also consistent with previous studies of multi-walled carbon nanotubes (J. Z. Liu et al., 2003; 2001), which modeled the multi-walled carbon nanotube (MWCNT) as a homogeneous continuum beam and successfully explained the rippling in MWCNT under bending in experiments (Poncharal et al., 1999). It should be noted that to model the post-buckling of the layered crystalline materials, such as the formation of kinking band, we may still require atomistic simulations (Li et al., 2007; Y. Liu et al., 2011), thin shell/plate FEA models (J. Z. Liu et al., 2005), or those atomistic-based FEA techniques (Arroyo and Belytschko, 2003).

There are many interesting layered crystalline materials that have other types of crystal symmetry (Geim and Grigorieva, 2013). For example, the perovskite-type material $LaNb_2O_7$, $(Ca,Sr)_2Nb_3O_{10}$, $Bi_4Ti_3O_{12}$ and $Ca_2Ta_2TiO_{10}$ have an orthorhombic symmetry. A theoretical model for an orthorhombic crystalline material is presented in Appendix. Note that for this symmetry, the column could undergo a mechanical buckling in either of the two lateral directions (*i.e.*, $y$ or $z$-axis in Fig. 1(b)), in comparison with only one direction (*i.e.*, $z$-axis) for the hexagonal crystalline materials. Consequently there is one more result for the IBS $\varepsilon_{cr}$. Please refer to Appendix for details.

From our theoretical model, we can conclude that the drastically low interlayer shear modulus $C_{44}$ in comparison with the in-plane modulus $C_{11}$ is the origin for the observed profound shear mode of instability and the anomalous elastic buckling of a column with an infinitesimal slenderness ratio (Fig. 1(a)). For a crystal material, the concept of elastic anisotropy degree $\delta(\boldsymbol{C})$ is often adopted to quantify the difference of elastic moduli along different crystalline directions (Nye, 1985; L.-F. Wang and Zheng, 2007). The special crystal structure of a layered material implies a high $\delta(\boldsymbol{C})$. Indeed, among the top 20 hexagonal crystal materials with a high $\delta(\boldsymbol{C})$, most of them are layered crystalline materials (L.-F. Wang and Zheng, 2007). It is natural to expect that a high elastic anisotropy degree $\delta(\boldsymbol{C})$ should lead to a small IBS $\varepsilon_{cr}$. However, a comparison between the IBS $\varepsilon_{cr}$ results and the anisotropy degree shows several exceptions. For instance, in Table I, $MoS_2$, Muscovite [$KAl_2Si_3O_{10}(OH,F)_2$], or Rubidium nickel chloride ($RbNiCl_3$) has a similar IBS $\varepsilon_{cr}$, *i.e.*, –0.1583, –0.1622, and –0.1552, but they show a significant difference in $\delta(\boldsymbol{C})$, *i.e.*, 0.608, 0.5, and 0.408. Another example is that although Muscovite [$KAl_2Si_3O_{10}(OH,F)_2$] has a larger $\delta(\boldsymbol{C})$ in comparison with Biotite [$K(Mg,Fe)_3AlSi_3O_{10}(OH,F)_2$], *i.e.*, 0.608 *vs.* 0.557, its IBS $\varepsilon_{cr}$ result is much higher as well, *i.e.*, –0.1583 *vs.* –0.0643. Here we propose that the IBS $\varepsilon_{cr}$ (Eq. (15)) could serve as an alternative measure to characterize the degree of elastic anisotropy for hexagonal crystal materials. In the same spirit, Eq. (A13) and (A14) could be used to measure the degree of elastic anisotropy for orthorhombic crystalline materials. One clear advantage is that such a measure, *i.e.* IBS $\varepsilon_{cr}$, has a clearer physical meaning.

In light of the very weak interlayer physical interactions, it is intuitively reasonable to approximate the critical buckling stress/strain of a multi-layered stack by that of a mono-crystal-layer (Guo et al., 2011). Our study shows that such an approximation might be problematic. For example, based on Euler model (Eq. (1)), for a graphene layer with length $L = 20$, 30, or 40 nm, the $\varepsilon_{cr}$ is –0.00358%, –0.00159%, and –0.000896% given the thickness of graphene monolayer as 0.066 nm (L. Wang et al., 2005; Yakobson et al., 1996), But in our MD simulations for a multi-layer stack, they share a similar $\varepsilon_{cr} = 0.0364\%$. It clearly shows that despite its small magnitude, the interlayer modulus $C_{44}$ plays a decisive role in determining the mechanical buckling of a multi-layered stack of graphenes.

Employing elastic buckling to tune the physical properties of layered crystalline materials has several clear advantages. First, there are no chemical or physical damages to the crystal integrity. It could avoid some undesired side effects that often occur when tailoring the physical properties via methods such as cutting, chemical or physical adsorptions. Second, in principle, the elastic buckling is recoverable. That means utilizing buckling under a cyclic loading/unloading condition can repeatedly control the material morphologies and thus their properties. This is highly desirable in nanotechnology, which can enable many new applications, such as the mechanical sensor, and the responsive materials.

Making use of the shear mode of instability has several more advantages. First, a layered crystalline material can undergo a mechanical buckling with a very low slenderness ratio ($L/\rho \sim 0$), suggesting the buckling wavelength $L$ can be tuned to a very small value. Our MD simulations demonstrate the scale of $L$ down to ~20 nm. It can be used to generate periodic surface structures at a nanometer scale, which is a difficult task by employing the bending mode of instability. Second, a distinctive kinking morphology is the signature of the shear mode instability at post-buckling stage (Budiansky et al., 1998; Z. Liu et al., 2010). In the kink, there is a sharp transition corner connecting two consecutive straight segments, which is potentially useful in some novel applications. For example, an electric current in graphene nano-bubbles can generate a giant pseudo-magnetic field (Levy et al., 2010). It was found that strength of the magnetic field depended on a change of curvature. The sharp corners in the kinks could be used to design nano-devices that can generate dynamically tunable giant pseudo-magnetic fields. Third, under the shear mode of instability, there are no strains in the basal planes of the layered crystalline materials (only shear deformation occurs among adjacent crystal layers). This could be another benefit, if the atomistic structure of crystal layers would like to be conserved.

## 6. Summary

In this paper, we study the elastic buckling of a column made of layered crystalline materials being subject to a uniaxial compressive load along the basal plane direction, using FEA simulation, MD simulations, and continuum mechanical modeling. FEA results show that with a reduction of the slenderness ratio $L/\rho$, there is a gradual transition from bending mode of instability to shear mode of instability. The effect of interlayer shear deformation appears to be much more significant than the isotropic materials. As the $L/\rho$ approaches to zero, the critical buckling strain $\varepsilon_{cr}$ converges to a value lower than mechanical strength of the materials, suggesting that the mechanical buckling should occur in the absence of structural slenderness. A continuum mechanics model is developed to understand this anomalous mechanical buckling. Our theoretical model reveals that the critical bucking strain $\varepsilon_{cr}$ at $L/\rho = 0$ solely depends on the material's elastic constants (with no structural dimensions), implying that it is an intrinsic material property. A new concept, intrinsic buckling strain (IBS), is thus defined. For a set of typical layered crystalline materials, theoretical results of IBS $\varepsilon_{cr}$ agree with FEA results very well. The good agreement with MD simulations for graphite and virtual graphite indicates that our model is applicable down to a nanometer scale (~20 nm). This theoretical model also reveals that a high degree of elastic anisotropy is the origin for the anomalous mechanical buckling in the absence of structural slenderness. Some in-depth discussions and potential applications in nanotechnology are provided. This study could provide guidelines for engineering layered crystalline materials in various nano-materials and nano-devices via mechanical buckling.

**Appendix A. Continuum mechanics model for intrinsic buckling strain of orthorhombic crystalline materials**

A continuum mechanics model is detailed here for a column made of orthorhombic crystalline materials (Fig. 1(b)). The mechanical buckling could occur in either of the two lateral directions. Thus periodic boundary conditions are applied in both $y$ and $z$ directions. Similar to the previous derivations in section 3 , boundary conditions are

$$
\begin{aligned}
&u_1\big|_{x=L/2} = -\delta/2, \quad \tau_{xy}\big|_{x=L/2} = 0, \quad \tau_{xz}\big|_{x=L/2} = 0 \\
&u_1\big|_{x=-L/2} = \delta/2, \quad \tau_{xy}\big|_{x=-L/2} = 0, \quad \tau_{xz}\big|_{x=-L/2} = 0 \\
&u_2\big|_{x=-L/2} = u_2\big|_{x=L/2}, \quad u_3\big|_{x=-L/2} = u_3\big|_{x=L/2}
\end{aligned}
\tag{A1}
$$

where $\delta$ is the relative displacement of the both ends moving toward each other. The displacement fields can be expressed as

$$u_1 = \varepsilon x + \sum_{n=1}^{N} a_n \frac{L}{2n\pi} \sin\frac{2n\pi x}{L}, \quad u_2 = -\varepsilon d_{12} y + \sum_{n=1}^{N} k_n \frac{L}{2n\pi} \cos\frac{2n\pi x}{L}, \quad u_3 = -\varepsilon d_{13} z + \sum_{n=1}^{N} b_n \frac{L}{2n\pi} \cos\frac{2n\pi x}{L}$$

(A2)

where $\varepsilon = \delta/L$ represents the homogeneous compressive strain in x-direction prior to elastic buckling and $d_{12} = C_{12}/C_{22}$, $d_{13} = C_{13}/C_{33}$ in which $C_{12}$, $C_{13}$, $C_{22}$ and $C_{33}$ are elastic constants. Note that $u_2$ has a different expression compared to the transverse isotropic materials, e.g., hexagonal crystalline materials. The first terms of $u_2$ and $u_3$ are adopted for a purpose of releasing normal stress $\sigma_y$ and $\sigma_z$ upon the compressive load $\delta$ in x direction. These displacement fields (Eq. (A2)) satisfy the boundary conditions (Eq. (A1)). Parameters $a_n$, $k_n$ and $b_n$ are unknown coefficients. Only when the load is above the critical buckling value $\delta_{cr}$, the $a_n$, $k_n$ and $b_n$ will have nonzero solutions.

Deformation gradient $F$ and the first Seth strain $E$ can be derived based on the displacement fields as

$$F_{11} = 1 + \frac{\partial u_1}{\partial x}, \quad F_{22} = 1 + \frac{\partial u_2}{\partial y}, \quad F_{33} = 1 + \frac{\partial u_3}{\partial z},$$
$$F_{12} = \frac{\partial u_1}{\partial y}, \quad F_{13} = \frac{\partial u_1}{\partial z}, \quad F_{23} = \frac{\partial u_2}{\partial z}, \quad \text{(A3)}$$
$$F_{21} = \frac{\partial u_2}{\partial x}, \quad F_{31} = \frac{\partial u_3}{\partial x}, \quad F_{32} = \frac{\partial u_3}{\partial y}$$

$$E_{11} = \left(F_{11}^2 + F_{21}^2 + F_{31}^2 - 1\right)/2, \quad E_{22} = \left(F_{12}^2 + F_{22}^2 + F_{32}^2 - 1\right)/2, \quad E_{33} = \left(F_{13}^2 + F_{23}^2 + F_{33}^2 - 1\right)/2$$
$$E_{12} = E_{21} = \left(F_{11}F_{12} + F_{21}F_{22} + F_{31}F_{32}\right)/2, \quad E_{13} = E_{31} = \left(F_{11}F_{13} + F_{21}F_{23} + F_{31}F_{33}\right)/2, \quad \text{(A4)}$$
$$E_{23} = E_{32} = \left(F_{12}F_{13} + F_{22}F_{23} + F_{32}F_{33}\right)/2$$

For an orthorhombic crystalline material, the constitutive law of linear elasticity can be expressed as

$$\begin{pmatrix} \sigma_x \\ \sigma_y \\ \sigma_z \\ \tau_{yz} \\ \tau_{zx} \\ \tau_{xy} \end{pmatrix} = \begin{pmatrix} C_{11} & C_{12} & C_{13} & & & \\ C_{12} & C_{22} & C_{23} & & & \\ C_{13} & C_{23} & C_{33} & & & \\ & & & C_{44} & & \\ & & & & C_{55} & \\ & & & & & C_{66} \end{pmatrix} \begin{pmatrix} \varepsilon_{xx} \\ \varepsilon_{yy} \\ \varepsilon_{zz} \\ \gamma_{yz} \\ \gamma_{zx} \\ \gamma_{xy} \end{pmatrix} \quad \text{(A5)}$$

where $\sigma$ and $\tau$ represent the normal and shear stress components, $\varepsilon$ and $\gamma$ denote the normal and shear strain components, and $C_{ij}$ are the stiffness constants. Substituting Eq. (A4) into Eq. (A5) yields the second Piola-Kirchhoff stress $T$ as

$$T_{11} = C_{11}E_{11} + C_{12}E_{22} + C_{13}E_{33}, \quad T_{22} = C_{12}E_{11} + C_{22}E_{22} + C_{23}E_{33},$$
$$T_{33} = C_{13}E_{11} + C_{23}E_{22} + C_{33}E_{33}, \quad T_{23} = 2C_{44}E_{23}, \quad T_{13} = 2C_{55}E_{13}, \quad T_{12} = 2C_{66}E_{12} \quad \text{(A6)}$$

Then the strain energy density $U$ is

$$U = 1/2(T_{11}E_{11} + T_{22}E_{22} + T_{33}E_{33} + 4T_{13}E_{13} + 4T_{23}E_{23} + 4T_{12}E_{12}) \tag{A7}$$

Integrating the energy density $U$ in the column leads to the potential energy as

$$W = \frac{1}{L}\int_{-L/2}^{L/2} U dx \tag{A8}$$

in which a unit length is taken along the $y$ and $z$ directions, owing to the periodic boundary conditions.

Following the principle of minimum total potential energy, partial derivatives of $W$ with respect to the undetermined coefficients $a_n$, $b_n$, and $k_n$ should be equal to zero. Thus,

$$\begin{aligned} g_1 &= \partial W / \partial b_1 = b_1\left(\left(C_{11} - 4C_{55}d_{13} - C_{12}d_{12} - C_{13}d_{13}\right)\varepsilon + 2C_{55}\right) + O(\varepsilon^2) = 0 \\ g_2 &= \partial W / \partial b_2 = b_2\left(\left(C_{11} - 4C_{55}d_{13} - C_{12}d_{12} - C_{13}d_{13}\right)\varepsilon + 2C_{55}\right) + O(\varepsilon^2) = 0 \\ &\ldots\ldots \\ m_1 &= \partial W / \partial k_1 = k_1\left(\left(C_{11} - 4C_{66}d_{12} - C_{12}d_{12} - C_{13}d_{13}\right)\varepsilon + 2C_{66}\right) + O(\varepsilon^2) = 0 \\ m_2 &= \partial W / \partial k_2 = k_2\left(\left(C_{11} - 4C_{66}d_{12} - C_{12}d_{12} - C_{13}d_{13}\right)\varepsilon + 2C_{66}\right) + O(\varepsilon^2) = 0 \\ &\ldots\ldots \\ f_1 &= \partial W / \partial a_1 = a_1\left(C_{11} + \left(3C_{11} - C_{12}d_{12} - C_{13}d_{13}\right)\varepsilon\right) + O(\varepsilon^2) = 0 \\ f_2 &= \partial W / \partial a_2 = a_2\left(C_{11} + \left(3C_{11} - C_{12}d_{12} - C_{13}d_{13}\right)\varepsilon\right) + O(\varepsilon^2) = 0 \\ &\ldots\ldots \end{aligned} \tag{A9}$$

where the quadratic and higher order terms of strain $\varepsilon$ are omitted. Note that derivatives with respect to $b_n$ ($n = 1, 2, 3, 4 \ldots$) always yields the same equation as

$$\left(C_{11} - C_{12}d_{12} - C_{13}d_{13} - 4G_{13}d_{13}\right)\varepsilon + 2C_{55} = 0 \tag{A10}$$

The derivatives with respect to $k_n$ ($n = 1, 2, 3, 4 \ldots$) leads to an equation as

$$\left(C_{11} - C_{12}d_{12} - C_{13}d_{13} - 4G_{12}d_{12}\right)\varepsilon + 2C_{66} = 0 \tag{A11}$$

The derivatives with respect to $a_n$ ($n = 1, 2, 3, 4 \ldots$) leads to an equation as

$$\left(3C_{11} - d_{12}C_{12} - d_{13}C_{13}\right)\varepsilon + C_{11} = 0 \tag{A12}$$

From Eq. (A10), (A11) or (A12), we obtained the IBS $\varepsilon_{cr}$ as

$$\varepsilon_{cr} = -\frac{2C_{55}}{C_{11} - C_{12}d_{12} - \left(4G_{13} + C_{13}\right)d_{13}} \tag{A13}$$

or

$$\varepsilon_{cr} = -\frac{2C_{66}}{C_{11} - C_{13}d_{13} - \left(4G_{12} + C_{12}\right)d_{12}} \tag{A14}$$

or

$$\varepsilon_{cr} = -\frac{C_{11}}{3C_{11} - C_{12}d_{12} - C_{13}d_{13}} \tag{A15}$$

Knowledge of the perovskite layered crystalline materials is quite limited (Geim and Grigorieva, 2013). A complete set of elastic constants for $LaNb_2O_7$, $(Ca,Sr)_2Nb_3O_{10}$, $Bi_4Ti_3O_{12}$, $Ca_2Ta_2TiO_{10}$ are not available. Therefore, a quantitative comparison between the theoretical results Eq. (A13) and (A14) and numerical simulations are not feasible at present.

Note that Eq. (A13) and Eq. (A15) can be reduced into Eq. (15) and (16) for the hexagonal crystalline materials, through letting $C_{22} = C_{11}$, $C_{23} = C_{13}$, and $C_{55} = C_{44}$ because of the hexagonal crystal symmetry and $d_{12} = 0$ due to the plane-strain conditions adopted in Eq. (15) and (16).

**Acknowledgement**

J. Z. Liu acknowledges the support of a seed grant from the engineering faculty at Monash University and acknowledges National Computational Infrastructure at Australian National University and Monash Sun Grid high-performance computing facility at Monash University for providing the computational resource. Q. S. Z. acknowledges the financial support from NSFC through Grant No. 10832005, the National Basic Research Program of China (Grant No. 2007CB936803), and the National 863 Project (Grant No. 2008AA03Z302).